\documentclass[letterpaper]{article}
\usepackage{aaai}
\usepackage{times}
\usepackage{helvet}
\usepackage{graphicx}
\usepackage{courier}
\usepackage{tabularx}
\usepackage{booktabs}
\usepackage{natbib}
\usepackage{subfig}
\usepackage{url}

\begin{document}
%

\title{Can Social Media Platforms Transcend Political Labels? An Analysis of Neutral Conservations on Truth Social}
\author{Chaitya Shah\textsuperscript{1, 5}, Ritesh Konka\textsuperscript{1, 5}, Gautam Malpani\textsuperscript{1, 5}, Swapneel Mehta\textsuperscript{2, 4, 5}, Lynnette Hui Xian Ng\textsuperscript{3}\\
\textsuperscript{1}Dwarkadas J. Sanghvi College Of Engineering, Mumbai, India, \\
\textsuperscript{2}Boston University, USA, 
\textsuperscript{3}Carnegie Mellon University, USA\\
\textsuperscript{4}Massachusetts Institute of Technology, USA,
\textsuperscript{5}SimPPL, USA\\
chaitya0623@gmail.com, riteshkonka999@gmail.com, gautammalpani33@gmail.com, swapneel@mit.edu, lynnetteng@cmu.edu
}

\maketitle
\begin{abstract}
\begin{quote}
There is a prevailing perception that content on a social media platform generally have the same political leaning. These platforms are often viewed as ideologically congruent entities, reflecting the majority opinion of their users; a prime example of this is Truth Social.
While this perception may exist, it is essential to verify the platform's credibility, acknowledging that such platforms contain meaningful insights with neutral stances. 
To this end, we examine the dissemination of Wikipedia links on the alt-right platform, Truth Social. 
Wikipedia is recognized for enforcing content neutrality and serves as a unique lens to analyze the objectivity of user-generated content on Truth Social. 
By scrutinizing Truths with and without Wikipedia links, identifying toxicity trends \& recognizing coordinated networks, we observe a lower level of  engagement and a tendency for Truths shared on Truth Social to cover more neutral topics when it includes Wikipedia links (Wiki Truths).
Given the significantly different engagement and nature of content shared of Wiki Truths against Non-Wiki Truths, we emphasize that we should not generalize the techno-political affiliation of a social media platform, but rather should investigate the content closely.

\end{quote}
\end{abstract}

\section{Background}
In February 2022, former US President Donald Trump introduced the Truth Social platform after his account was suspended from Twitter. 
Truth Social allows users to create posts termed as ``Truths" that have text and media, and interact with other content using favourites, replies and retruths (reposts). 
Truth Social joins the set of social media platforms known as ``alt-tech" that includes Parler and Gab, labeled to serve users associated with the alt-right political movement \citep{dehghan2022politicization}. 
However, while a platform is labeled as ``alt-right", are the conversations within it consistently aligned with right-wing ideologies? Or could there be space for neutral Truths? How do the platform audiences respond when such links are shared within it?
To answer these questions, we look to investigate the sharing of Wikipedia links on Truth Social. 
Wikipedia commits to providing objective information and maintaining itself as a fair and impartial platform \citep{greenstein2012wikipedia}. 
We assume that Truths that share Wikipedia links are generally neutral, serving primarily to disseminate knowledge, because Wikipedia links are generally not associated with a partisan meaning and has factual backing.
The scholarly works by \citet{flisfeder2018trump}, \citet{altrightmemes} and \citet{neiwert2017alt} provide compelling arguments suggesting that the primary beneficiary of the 2016 US presidential elections was, assisted by an extreme right-wing community, the ``alt-right". 
Consequently, the establishment of the Truth Social platform was anticipated to serve as a sustaining force to this community, providing a digital space conducive to their ideologies and facilitating their continued influence in the socio-political landscape.

\citet{sharot2020people} showcase that the importance of social media on people's intake of information has gained significant attention in recent times. 
The platform’s distinct qualities, driven by hard-line supporters of the former president, position it as an influential mouthpiece, yet little is known about it \citep{gerard2023truth}. 
Thus, the way attention was brought to the platform, we aim to analyse if the entire Truth Social community is truly polarized? 
Or are there a group of topics where neutral discussions occur?

In order to identify such a group, we need to target a set of truths that share neutral nuances in general, and then check whether they perform similarly or become biased inside the platform. 
\citet{neutralwiki} claims that Wikipedia shares a "neutral point of view" (NPOV), but it is the multiple ambiguous voices that contribute to the shape of neutrality on the platform. 
Thus, with the help of these voices captured on Truth Social, we aim to capture the objectivity of the audience and discern how it may respond to different interactions within the platform.

In this paper, we address the Issue of Principle when studying social media platforms. 
In particular, social media platforms are often labelled with affiliations due to the alliances of their owners as stated in articles \citet{rsf2022} and \citet{pewresearch_truthsocial}. 
We question this generic labeling, for there are techno-political nuances within the content and the users of the platforms. 
While Truth Social is termed as an alt-tech, past analysis shows news reports of Truths, the stories shared on the platform reflect a wide range of content placed across the political spectrum \citep{zhang2024trump}.

This paper builds on past work on the analysis of the Truth Social (TS) (\citet{wilber2017trump}) and Wikipedia (Wiki) platforms (\citet{ortega2012wikipedia}, \citet{brandes2009network}), identifying the perceptions of content from one platform shared on another.
Using a combination of statistical and network analysis, we demonstrate that Truths with neutral Wikipedia links exhibit varying levels of engagement compared to Truths without such links. 
The coordination that exists between these communities showcases a higher engagement on alt-right topics in Truths without wiki-links, whereas Truths featuring wiki-links fosters more discussion oriented interactions. 
Our work underscores the importance of investigating this group of Truths.

\section{Retrieving Wikipedia Links on Truth Social}
We collected Truths from Truth Social using a custom tool based on Stanford's open-source system\footnote{\texttt{\url{https://github.com/stanfordio/truthbrush}}} to collect posts on the dates August 17, 2023 and December 5, 2023. 
This helped us collect the content of the Truths along with their creation time, author account details, media attachments, tags, counts of the number of favourites, replies and retruths.
We separated this data manually into Wiki Truths, identified by the presence of ``en.wikipedia.org" in the post, and Non-Wiki Truths. 
This results in a total of 1,790,749 Truths, of which there are 17451 unique Wiki Truths, or 1\% of the dataset, among 15313 unique users. 
Within these, we found out that there were 13831 unique links shared. 
There were 14503 truths containing only one Wikipedia link, 246 truths containing more than one Wikipedia link, and 2138 truths containing a Wikipedia link along with atleast one other non-Wikipedia link. 

\begin{figure}[htp]
    \includegraphics[width=0.5\textwidth]{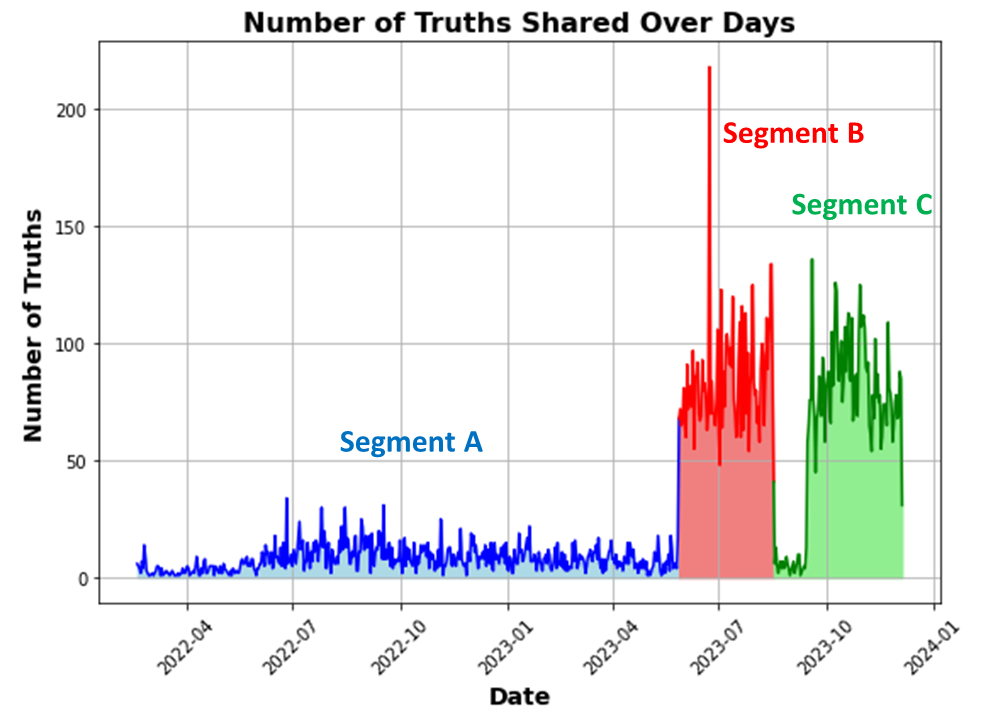}
    \caption{Count for Truths with Wikipedia Links across time}
    \label{fig:galaxy}
\end{figure}

Our analysis delineates the dataset into three distinct segments illustrated in Figure \ref{fig:galaxy}. 
Segment A, preceding a significant spike on May 28, 2023, is marked by \citet{cnbc2023} reports of Truth Social facing financial losses and a potential shutdown, prompting increased sharing of content from multiple accounts. \citet{economicTimes2023} also states the platform losing \$73 Million since it's launch. 
The Digital World Acquisition Corp states that a potential merger has become a necessity for Trump Media \& Technology Group, the owner of Truth Social.
August 17, 2023, witnesses Trump's return to X after a prolonged absence, diverting attention from Truth Social, resulting in Segment B coinciding with a dip. 
Since then, his account on X has been inactive. However, he also expressed his affection for Truth Social, stating that it is his home directly on the platform \citep{apnews2023}.
Segment C suggests a period of reevaluation and adaptation for the platform, contending with competition from Trump's presence on other platforms. 
By segmenting the dataset in this manner, we aim to gain insights, keeping an ever-evolving social media landscape into consideration.

\section{Exploratory Data Analysis}

\begin{table*}[]
  \centering
  \begin{tabular}{|c|c|c|c|} 
    \hline
    \textbf{Engagement Statistics} & \textbf{Segment A (wiki/non-wiki)} & \textbf{Segment B (wiki/non-wiki)} & \textbf{Segment C (wiki/non-wiki)} \\
    \hline
    favourites count & 1.90 $\pm$ 13.84 / \newline 6.83 $\pm$ 159.11 ** & 1.22 $\pm$ 13.31 / \newline 4.84 $\pm$ 108.72 **
& 1.35 $\pm$ 7.43 / \newline 8.68 $\pm$ 212.40 **\\
    \hline
    retruths count & 0.71 $\pm$ 5.01/ \newline 2.59 $\pm$ 47.39 *& 0.52 $\pm$ 5.49 / \newline 1.79 $\pm$ 33.14 **& 0.74 $\pm$ 4.21 / \newline 2.91 $\pm$ 63.38 **\\
    \hline
    replies count & 0.52 $\pm$ 1.28 / \newline 0.97 $\pm$ 45.07 **
& 0.52 $\pm$ 2.28 / \newline 0.54 $\pm$ 8.99 **
& 0.52 $\pm$ 1.39 /  \newline  0.98 $\pm$ 45.08 **\\
    \hline
    average toxicity & 0.12$\pm$0.15 / \newline 0.12$\pm$0.13  & 0.15$\pm$0.17 / \newline 0.11$\pm$0.12 & 0.14 $\pm$0.16 / \newline 0.14$\pm$ 0.14 \\
    \hline
  \end{tabular}
  
  \caption{Comparison of engagement statistics of Truths with and without Wikipedia Links. 
  ``*'' means the engagement between the two groups of Truths are significantly different at $p<0.05$, ``**'' is at the $p<0.01$ level.}
  \label{tab:engagement}
\end{table*}

Metrics for user engagement and content interaction across social media platforms were analyzed between the Wiki/Non-Wiki Truths using the Kruskal-Wallis test (Table \ref{tab:engagement}). 
This test was chosen due to its non-parametric nature, suitable for analyzing data that does not adhere to the normal distribution. 
Wiki Truths consistently exhibited significantly lower engagement metrics across all time segments, suggesting reduced user interaction compared to non-Wiki Truths. 
The analysis revealed consistent levels of neutrality and objectivity associated with Wikipedia content, as evidenced by stable, lower toxicity scores across segments.
The findings underscore that within politically right-leaning platforms there are "pockets of neutrality" exhibiting behaviors that depart from platform norms, potentially related to similar findings about scientifically organized pockets within COVID-skeptic communities \citep{lee2021viral}.
Truths were labeled with toxicity values using the Perspective API\footnote{https://perspectiveapi.com/}. 
Both Wiki and Non-Wiki Truths generally have a low average toxicity score, which suggests there are nuances of alt-right platforms that are overlooked by current literature (e.g. toxicity analysis) and may warrant a longitudinal study \citep{rieger2021assessing}. There may be other subtle linguistic differences between Wiki and Non-Wiki Truths, that suggests that deeper analysis into the evaluation of content in alt-right platform might require specialized classifiers to identify topics and language of interest.

\begin{figure*}[btp]
    \centering
    \includegraphics[width=\textwidth]{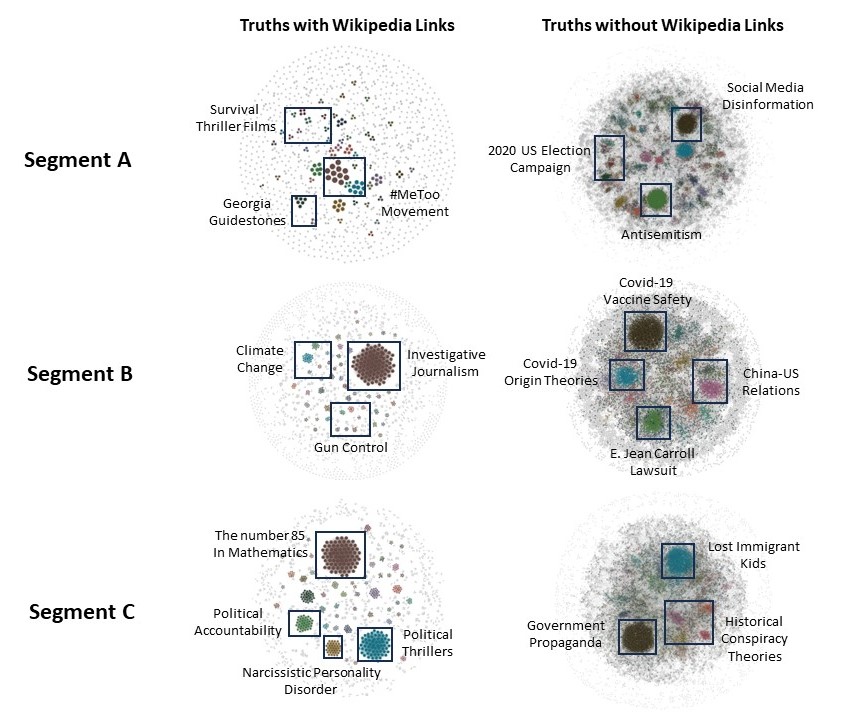}
    \caption{Coordination Graphs by Common Links for Truths Containing Wikipedia and Non-Wikipedia Links.}
    \label{fig:galaxy}
\end{figure*}

\section{Coordinated Sharing of Wikipedia Links}
We then investigated coordinated link sharing behaviors on Truth Social, focusing on users sharing the same Wikipedia link. 
Using temporal coordination identification techniques \citep{ng2022online}, we constructed coordination network graphs across each time segment. The time factor for analyzing the coordination was each entire segment.
In these graphs, nodes represent users, and edges between two nodes indicate that they shared the same Wikipedia link within the respective time segment. 

In Segment A, minimal coordination was observed, characterized by numerous small clusters. 
While Segment B showed a dominant large cluster along with smaller clusters, Segment C demonstrated even more pronounced coordination, with three to four significant clusters alongside smaller ones, yielding a higher graph density. 
This variance in the presence of coordinated link sharing across Segments A through C imply a shift in user behavior. 
Moreover, the presence of coordination among users sharing neutral links in Segments B and C suggests that not all coordination is inherently malicious. 
The transient nature of coordinated clusters emphasizes the potential for co-sharing behaviors to occur.


\section{Coordinated Sharing of Non-Wikipedia Links}
The analysis of non-Wikipedia data reveals significant coordination across all three segments, with all clusters showing notable coordination of non-Wikipedia links. Given the inherently larger volume of Truths within the non-wiki dataset compared to the Wikipedia dataset, it is unsurprising to observe denser graphs for non-Wikipedia Data. This observation underscores the platform's active engagement in coordinated link-sharing activities for non-Wikipedia topics, indicative of the community's interest in such content. 

We manually sieved through the Truths in the same cluster and provided labels of the topics. The topics extracted in our study provided insights into prevalent themes discussed within each community. Within the Non-Wiki Truths, prominent topics included "Lost Immigrant Children", "Government Propaganda", "COVID-19 Vaccine Safety"", and "Political Conspiracy Theories", reflecting a focus on politically charged issues and societal concerns. Conversely, Wiki Truths featured a broader range of topics, encompassing "Survival Thriller Films", "Investigative Journalism", "climate Change", and "Mathematical Concepts". This diversity suggested a neutral stance towards information sharing within the Wikipedia community, highlighting the potential for fostering balanced discourse and informed discussions on Truth Social. The comparative analysis of topics extracted from Wiki and Non-Wiki underscores the importance of studying neutral groups within alt-right platforms. While Non-wiki clusters predominantly focused on politically charged issues and conspiracy theories, Wiki clusters covered a broader spectrum of subjects with a neutral stance. These findings offer valuable insights into the diverse interests and engagement patterns within these communities, informing efforts to promote informed discussions and balanced discourse on Truth Social.

\section{Discussion}
Studying alt-right platforms helps in analyzing disinformation campaigns \citep{bevensee2018alt}. Continuous study of such social media platforms is essential for effectively addressing the impact of extremist movements on society and fosters a deeper understanding of societal vulnerabilities and resilience strategies \citep{zannettou2018gab}. Studying the Truths with Wikipedia links on such platforms may indicate topics and issues significant within the community of users that share the same ideology. This allows understanding of user interests which can provide insights into the ideological priorities and concerns of alt-right groups \citep{chadwick2022amplification}.





While our analysis offers valuable insights into the engagement dynamics on Truth Social, it is important to acknowledge the inherent limitations of our study. Search results on Truth Social gives us only a maximum of 10,000 Truths, therefore creating a data accessibility problem. By searching the platform multiple times, we are able to collect a larger set of Truths. Our method is also prone to sampling bias, for the returned set of Truths likely contained the most popular Truths on the platform. However, given that the most popular Truths are the most viewed and recommended Truths, we believe that this set sufficiently represents the platform.

The analysis of engagement metrics between Wiki and Non-Wiki Truths, conducted across three distinct time segments, revealed notable disparities in user interaction on Truth Social. Notably, Wiki Truths consistently exhibited lower engagement metrics across all time segments compared to Non-Wiki Truths, indicating reduced interaction. Furthermore, the analysis highlighted consistent levels of neutrality and objectivity as evidenced by stable toxicity scores across segments. In addition to quantitative metrics, manual labeling of topics within Truths clusters provided insights into prevalent themes discussed within each community. Non-Wiki Truths Clusters primarily addressed politically charged topics and conspiracy theories, whereas those associated with Wiki Truths encompassed a wider range of subjects, reflecting a more neutral stance.
These findings underscore the importance of studying neutral groups within alt-right platforms like Truth Social, offering valuable insights into the diverse interests and engagement patterns within these communities.

Given the differences in engagement statistics and the transient coordinated link sharing behavior between Wiki/ Non-Wiki links, we urge the computational social science community to not judge a platform by its label. This work draws into the topic of theme of the issue of principle on social media platforms. Very often, platforms are stereotyped with political alliances due to their owners. These techno-political alliances undoubtedly create a filter for social computing studies. Studies on the Truth Social platforms typically revolve around its alt-right characteristics, such as content analysis reflecting the support of the Republican party, or investigations of conspiracy-laden narratives \citep{gerard2023truth}. Neither the creator nor the owner of the platform should entirely determine the political affiliation of a platform. Instead, the users themselves, and the conversations within determine the platform's character. In that aspect, we put forth that every platform has a silver lining: the objective, innocuous, non-toxic posts. Therefore, such social media platforms should not be judged by their political labels, but  we should study the platform by itself to understand its character.

\section{Conclusion} 

The analysis conducted explored the engagement of the content shared on Truth Social. Wiki Truths have a significantly less engagement, which might be because they are neutral Truths and thus do not illicit much interactions. The study also focuses on coordinated link sharing on Truth Social, extracting topics from clusters identified in both Wikipedia (wiki) and non-Wikipedia (non-wiki) Truths. Wiki and Non-Wiki Truths have different properties in terms of engagement statistics, coordination patterns and topics shared, emphasizing that the content on a social media platform transcends political labels and should not be labeled with a sweeping statement. This work illustrates the importance of scrutinizing different categories of content within a social media platform, and opens discussion for future research into the characteristics of sub-populations of political communities on social media.

\bibliographystyle{aaai}

\end{document}